\documentclass[aps,pra,showpacs,twocolumn,superscriptaddress]{revtex4-1}

\usepackage{hyperref}
\usepackage{graphicx}
\usepackage{amsmath,amssymb,latexsym,amsfonts,subfigure,bbm}
\usepackage{dsfont,relsize}
\begin{document}

\title{Scattering-free pulse propagation through invisible non-Hermitian disorder}

\author{Andre Brandst\"otter}
\affiliation{Institute for Theoretical Physics, Vienna University of Technology (TU Wien), 1040, Vienna, Austria, EU}

\author{Konstantinos G.~Makris}
\affiliation{Department of Physics, University of Crete, 71003, Heraklion, Greece, EU}

\author{Stefan Rotter}
\affiliation{Institute for Theoretical Physics, Vienna University of Technology (TU Wien), 1040, Vienna, Austria, EU}

\begin{abstract}
We demonstrate a new design principle for unidirectionally invisible non-Hermitian structures that are not only invisible for one specific wavelength but rather for a broad frequency range. Our idea is based on the concept of constant-intensity waves, which can propagate even through highly disordered media without back-scattering or intensity variations. Contrary to already existing invisibility studies, our new design principle requires neither a specific symmetry (like $\mathcal{PT}$-symmetry) nor periodicity, and can thus be applied in a much wider context. This generality combined with broadband frequency stability allows a pulse to propagate through a disordered medium as if the medium was entirely uniform.

\end{abstract}

\maketitle

The idea to confer new properties on materials by adding an appropriate distribution of gain and loss to them has recently received considerable attention \cite{feng_non-hermitian_2017, el-ganainy_non-hermitian_2018, kottos_optical_2010, longhi_parity-time_2017}. At first glance one may expect that mixing loss with gain just results in a mutual cancellation of the effects of these components. The physics observed in such scenarios is, however, very rich and full of surprises. One particular case that has been studied extensively is that of synthetic materials obeying a so-called parity-time ($\mathcal{PT}$) symmetry \cite{bender_real_1998, bender_complex_2002}. In the framework of photonics, $\mathcal{PT}$-symmetry has served as a new design principle for engineering composite structures with gain and loss \cite{feng_non-hermitian_2017, el-ganainy_non-hermitian_2018, longhi_parity-time_2017, ruter_observation_2010, kottos_optical_2010, guo_observation_2009, makris_beam_2008, el-ganainy_theory_2007, musslimani_optical_2008} that feature a plethora of remarkable characteristics like power oscillations \cite{makris_beam_2008, klaiman_visualization_2008}, non-reciprocal transport \cite{nazari_optical_2014, peng_parity-time-symmetric_2014, chang_parity-time_2014}, loss-induced transparency \cite{guo_observation_2009}, perfect absorption \cite{longhi_pt-symmetric_2010, chong_pt_2011, sun_experimental_2014}, and loss-induced lasing \cite{liertzer_pump-induced_2012, brandstetter_reversing_2014, peng_loss-induced_2014}.

One of the most successful concepts that has emerged from the field of $\mathcal{PT}$-symmetric optics so far is the idea to make periodic gratings unidirectionally invisible by adding loss and gain to them in a well-controlled fashion \cite{lin_unidirectional_2011}. In such systems, the reflection from one end of the structure is zero while it is increased from the other end. Moreover, the transmission from both sides is perfect and the accumulated phases along the two propagation directions are the same as in the absence of the structure. The first theoretical proposal for such unidirectionally invisible structures \citep{lin_unidirectional_2011} drew considerable attention and was successfully implemented by several experimental groups \cite{regensburger_parity-time_2012, feng_experimental_2013, feng_demonstration_2014}. The idea was later extended to non-$\mathcal{PT}$-symmetric potentials, which, however, are restricted to layered systems \cite{mostafazadeh_invisibility_2013}, periodic systems \cite{longhi_half-spectral_2015} or which have to be analytic in one half of the complex position plane (in terms of spatial Kramers-Kronig relations) \cite{horsley_spatial_2015, longhi_wave_2015, horsley_wave_2016}. In spite of the intense research activities related to this novel concept, the question whether this concept can also be generalized to aperiodic, non-$\mathcal{PT}$-symmetric and non-analytic potentials remains to be answered.

Here, we propose such a general design principle for unidirectionally invisible structures that are unrestricted in their spatial shape. In fact, we can even make \textit{disordered} structures unidirectionally invisible, which give rise to strong scattering \cite{lagendijk_resonant_1996, akkermans_mesoscopic_2007, sebbah_waves_2001, rotter_light_2017, mosk_controlling_2012} or even Anderson localization \cite{anderson_absence_1958, wiersma_localization_1997, lagendijk_fifty_2009, segev_anderson_2013}. Our approach works not only for one specific wavelength but rather for a broad frequency range. In this way a pulse can propagate through the disordered medium like through uniform space. Moreover, we show that the unidirectional invisible Bragg grating proposed in Ref.~\cite{lin_unidirectional_2011} coincides with one example of our new family of invisible structures at specific parameter configurations. When moving away from these parameter values the $\mathcal{PT}$-symmetric Bragg grating loses its invisibility property, whereas our new system stays invisible. 
    
The design principle we introduce here for creating unidirectionally invisible structures is based on the concept of so-called constant-intensity (CI) waves \cite{makris_constant-intensity_2015, makris_wave_2017, yu_bohmian_2018} that has recently been realized also experimentally with an acoustic setup \cite{rivet_constant-pressure_2018}. These CI waves have the remarkable feature that they are perfectly transmitted even through strongly disordered structures \cite{makris_wave_2017}. The only signature that they carry from the disordered potential is an extra phase as compared to propagation through free space. Since a phase measurement can thus reveal the presence of such CI scattering potentials, they are in general not invisible (only perfectly transmitting). Here, we present a way to eliminate the possibility of detecting such potentials even through phase measurements. Moreover, we show that the invisibility obtained with our approach is a broadband feature that can even be used for sending pulses across these potentials which maintain their spatial profile during the entire transmission process.

\begin{figure}[t]
\includegraphics[clip,width=1.0\linewidth]{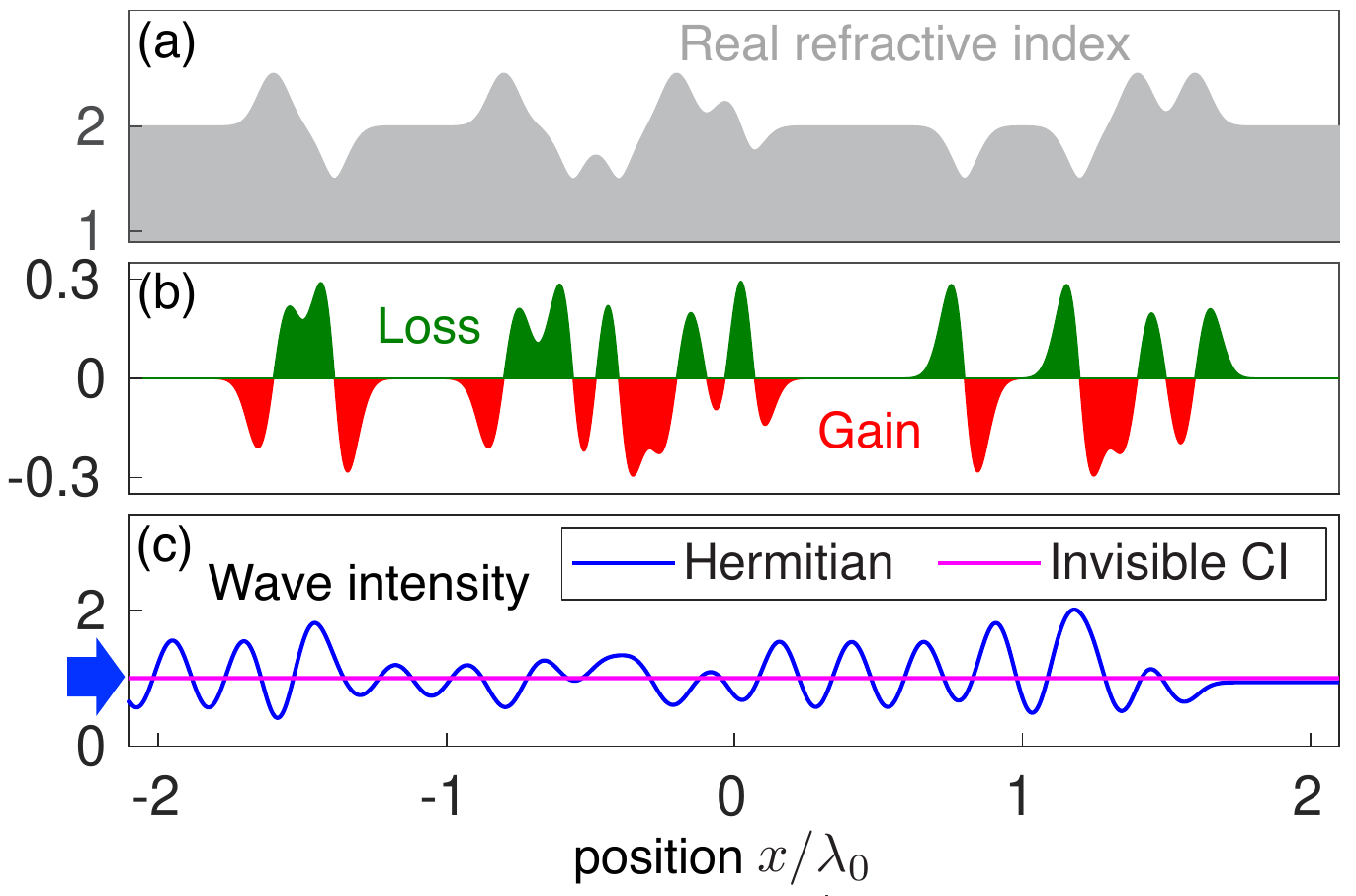}
\caption{(a), (b) Real (gray) and imaginary part (red: gain, green: loss) of the refractive index for a CI system that is invisible from the left. The system follows Eq.~(\ref{eq:CI_ref_index}) with $W(x) = n_0+f(x)$, where $f(x)$ is the superposition of twelve randomly placed Gaussians, with six of them having a positive and six of them a negative amplitude as is dictated by Eq.~(\ref{eq:invisibility_condition}). The asymptotic refractive index is $n_0=2$ and the wavenumber $k_0=2\pi/0.2$. (c) Intensity of the scattering state for a plane wave with wavenumber $k_0$ and amplitude $A=1$ injected from the left into the system for the structure including gain and loss (magenta line) and without gain and loss (blue line). As can be seen, the non-Hermitian components make the wave perfectly transmitting without any intensity variations.}\label{fig:1}
\end{figure} 

Starting point for our investigation is the Helmholtz equation that describes the stationary scattering state of a linearly polarized electric field $\psi(x)$ in a one-dimensional slab, 
\begin{equation}
\label{eq:helmholtz}
[\partial_x^2 + n^2(x)k_0^2]\psi(x)=0.
\end{equation}
Here $n(x)=n_R(x)+\mathrm{i}n_I(x)$ is the complex refractive index and $k_0=2\pi/\lambda_0$ is the vacuum wavenumber with $\lambda_0$ being the vacuum wavelength. $n_R(x)$ describes the real index variations whereas $n_I(x)$ describes the gain-loss profile. One such one-dimensional refractive index distribution is shown in Fig.~\ref{fig:1}, where $n_R(x)$ is depicted in gray [see Fig.~\ref{fig:1}(a)] and $n_I(x)$ in red (gain) and green (loss) [see Fig.~\ref{fig:1}(b)]. In general, when an incident plane wave propagates through a spatially varying distribution $n(x)$, the interference between forward and backward propagating waves leads to a complex intensity pattern. These intensity variations can, however, be entirely eliminated in a certain class \cite{makris_wave_2017, makris_constant-intensity_2015} of non-uniform complex index distributions \cite{wadati_construction_2008, konotop_families_2014, tsoy_stable_2014, nixon_all-real_2016}, resulting in a perfectly transmitted wave with the same constant intensity (CI) in all of space. To observe these special wave states, the real and imaginary part of the refractive index have to satisfy
\begin{equation}
\label{eq:CI_ref_index}
n^2(x)= [n_R(x)+\mathrm{i}n_I(x)]^2=W^2(x)-\frac{\mathrm{i}}{k_0}\partial_x W(x),
\end{equation}
with $W(x)$ being an arbitrary real-valued function related physically to the Poynting vector power flow \cite{makris_constant-intensity_2015}. The CI solution of the Helmholtz Eq.~(\ref{eq:helmholtz}) with a refractive index in Eq.~(\ref{eq:CI_ref_index}) is a right-propagating wave $\psi(x)=A\, e^{\mathrm{i}k_0\int_{-L}^x W(x') dx'}$, where $-L$ and $L$ are the left and right borders of the scattering region of width $2L$ and $A$ is a constant amplitude. The unique feature of this solution is that it has a constant intensity inside the scattering region, i.e., $I=|\psi(x)|^2=|A|^2$, despite the fact that the index of refraction in Eq.~(\ref{eq:CI_ref_index}) is non-uniform. The real-valued function $W(x)$ can be chosen arbitrarily and serves as a ``generating" function to produce CI refractive indices via Eq.~(\ref{eq:CI_ref_index}). Radiation boundary conditions of the electric field that assure perfect transmission impose the following condition for $W(x): W(-L)=W(L)=n_0$, where $n_0$ is the refractive index of the asymptotic regions. In this way a plane wave with wavenumber $k_0$  incident from the left asymptotic region $x<-L$ will feature a constant intensity inside the non-uniform scattering region between $-L<x<L$. While it will also be perfectly transmitted to the right asymptotic region $x>L$, the scattering potential still imprints information on its shape onto the transmission phase $\phi_t = k_0\int_{-L}^L W(x)dx$ of the outgoing plane wave.

Here we show that we can tune this transmission phase $\phi_t$ in such a way that the outgoing plane wave carries no information on the scattering region at all. In other words, we demonstrate how to make scattering potentials as described by Eq.~(\ref{eq:CI_ref_index}) unidirectionally invisible. We start by choosing the generating function to be of the form 
\begin{equation}
\label{eq:generating_function}
W(x)=n_0 + f(x),
\end{equation}
where $n_0$ is the refractive index of the asymptotic regions $|x|>L$ and $f(x)$ is an arbitrary real-valued function that should satisfy 
\begin{equation}
\label{eq:invisibility_condition}
\int_{-L}^L f(x) dx = 0.
\end{equation}
This function $f(x)$ describes the phase the CI wave accumulates additionally to the propagation through a uniform medium with index $n_0$. Thus, enforcing Eq.~(\ref{eq:invisibility_condition}) is equivalent to demanding that this additional phase vanishes. To be more precise, the transmission phase of a CI wave with a generating function satisfying Eq.~(\ref{eq:invisibility_condition}) takes the value $\phi_t = k_0\int_{-L}^L [n_0+f(x)]dx=2 k_0 L n_0$ which is equal to the phase a wave would accumulate by propagating through a scattering region of width $2L$ with the same uniform index of refraction $n_0$ as in the asymptotic regions. Neither the transmitted intensity nor the transmitted phase then reveal whether the refractive index is uniform with $n_0$ or an inhomogeneous refractive index distribution. As we will show explicitly below, the broadband stability of the CI waves we create in this way naturally also gives rises to the same time-delay $\tau=d \phi_t/dk$ as obtained in the uniform system, not only at the target frequency $k_0$ but rather in a sizable frequency window. In other words, even time-resolved measurements on wave-packets closely centered around the design frequency $k_0$ will not be able to detect the presence of the unidirectionally invisible medium we propose here. 

\begin{figure}[t]
\includegraphics[clip,width=1\linewidth]{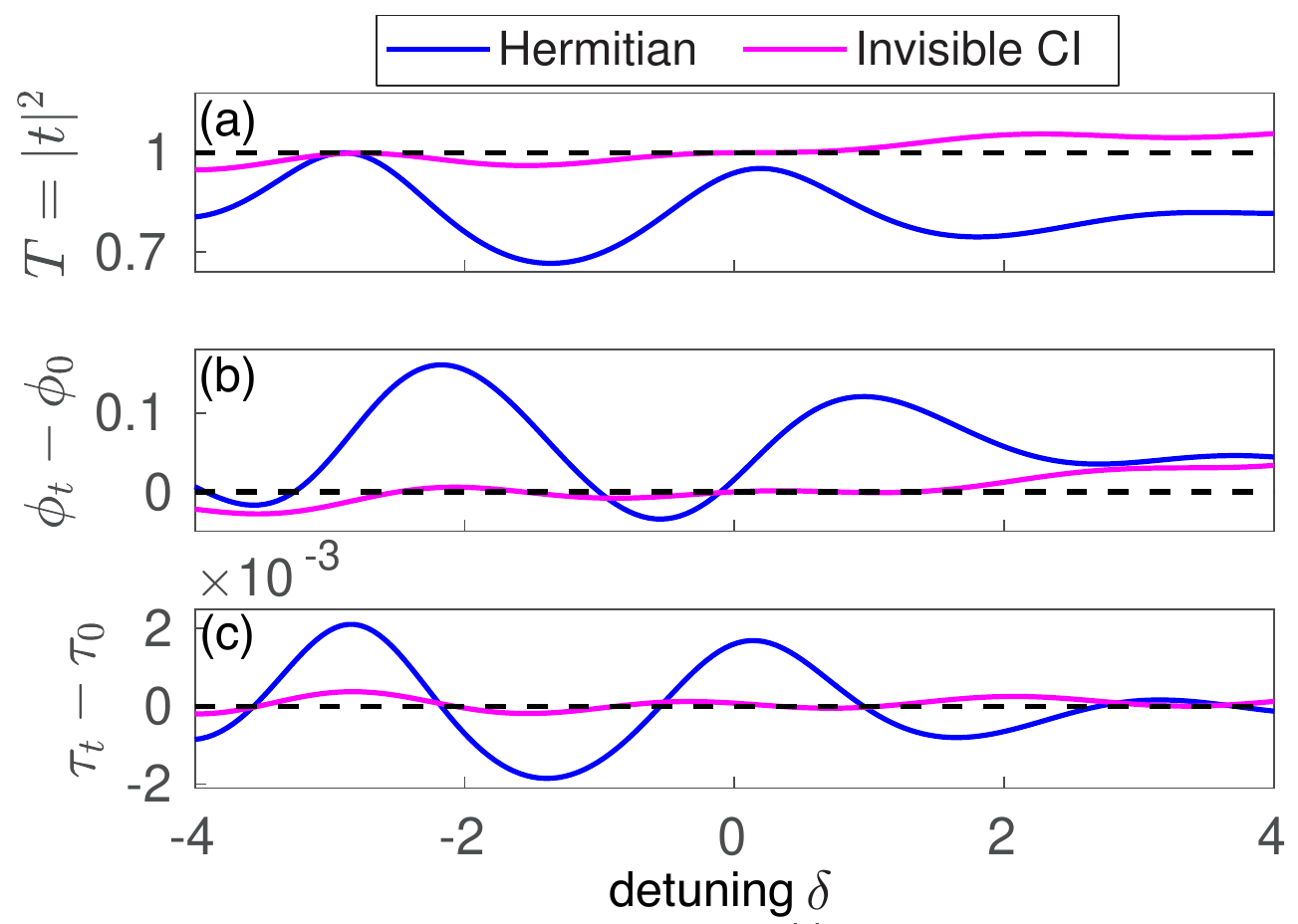}
\caption{(a) Transmission spectrum of the Hermitian (blue line) and invisible CI system (magenta line) shown in Fig.~\ref{fig:1} as a function of the wavenumber detuning $\delta=k-k_0$. (b) Phase of the transmission amplitude $\phi_t$ minus the phase that the wave would accumulate in a uniform material $\phi_0=2 L k n_0$ for the Hermitian and for the invisible CI system. (c) Difference of the corresponding time-delays $\tau_t - \tau_0 = d\phi_t/dk-d\phi_0/dk$. For the invisible CI system the transmittance in (a) is close to unity and the other two quantities in (b) and (c) are close to zero in a broad interval. The reference values of perfect transmittance, zero phase difference and zero time-delay as necessary for a system to be perfectly invisible are indicated by the horizontal dashed lines. The small deviations from perfect invisibility (see horizontal dashed lines) turn out to be around two orders of magnitude smaller for the CI system than for the Hermitian system (see appendix \ref{AppA}). The relative width of the invisibility window between $\delta=-2$ and $\delta=2$ is $\Delta \delta / k_0 \approx 4/31.42 \approx 0.13$, i.e., a wavenumber detuning of around $6\%$ from $k_0$ in both directions still allows for perfect transmission and zero accumulated phase.}\label{fig:2}
\end{figure} 

In order to test these predictions and their broad applicability, we now investigate several interesting examples numerically. Consider first the refractive index distribution as provided in Eq.~(\ref{eq:generating_function}), where $f(x)$ consists of twelve randomly placed Gaussians with the same height and the same width but with six of them having a positive amplitude and six of them having a negative amplitude, thus enforcing the condition shown in Eq.~(\ref{eq:invisibility_condition}). The corresponding complex refractive index calculated with Eq.~(\ref{eq:CI_ref_index}) is shown in Fig.~\ref{fig:1}(a) and (b). To highlight that our theory does not rely on a smooth potential (as necessary for applying a semi-classical approximation), we consider here the case where the wavelength $\lambda_0=2\pi/k_0$ is larger than the variations of the refractive index. In Fig.~\ref{fig:1}(c) we display the intensity of the scattering state at the target frequency $k_0$ for the two cases with the gain and loss distribution added (magenta line) and without it (blue line). We can clearly see that the wave's intensity shows strong variations in the Hermitian case, whereas the intensity is constant for the system including gain and loss. The next step is to show that this CI system features unidirectional invisibility. In Fig.~\ref{fig:2}(a) we show the transmission spectrum, i.e., the transmittance $T=|t|^2$ as a function of the wavenumber detuning $\delta = k-k_0$, for the Hermitian (blue line) and for the CI system (magenta line) shown in Fig.~\ref{fig:1}. We see first of all that the CI system is close to perfectly transmitting not only at $k=k_0$ (i.e., $\delta=0$) but also in a broad frequency range around $k_0$ (between $\delta=-2$ and $\delta=2$), whereas the Hermitian system strongly deviates from unit transmittance. Fig.~\ref{fig:2}(b) shows the difference between the transmission phase $\phi_t$ of a wave propagating through the Hermitian or through the CI system as compared to the transmission phase of a wave propagating through a uniform material, $\phi_0=2 L k n_0$. We see that for the CI system the difference is close to zero in a broad frequency interval. In Fig.~\ref{fig:2}(c) we also show the difference between the corresponding time-delay $\tau_t$ as compared to the time-delay of a wave propagation through a uniform material $\tau_0 = d \phi_0/dk$. Also here the CI system yields the same values as the corresponding uniform system. Fig.~\ref{fig:2} thus clearly shows that the CI system in Fig.~\ref{fig:1} cannot be distinguished from a uniform system, i.e., it is indeed invisible from the left around the target frequency $k_0$.

\begin{figure}[t]
\includegraphics[clip,width=1.0\linewidth]{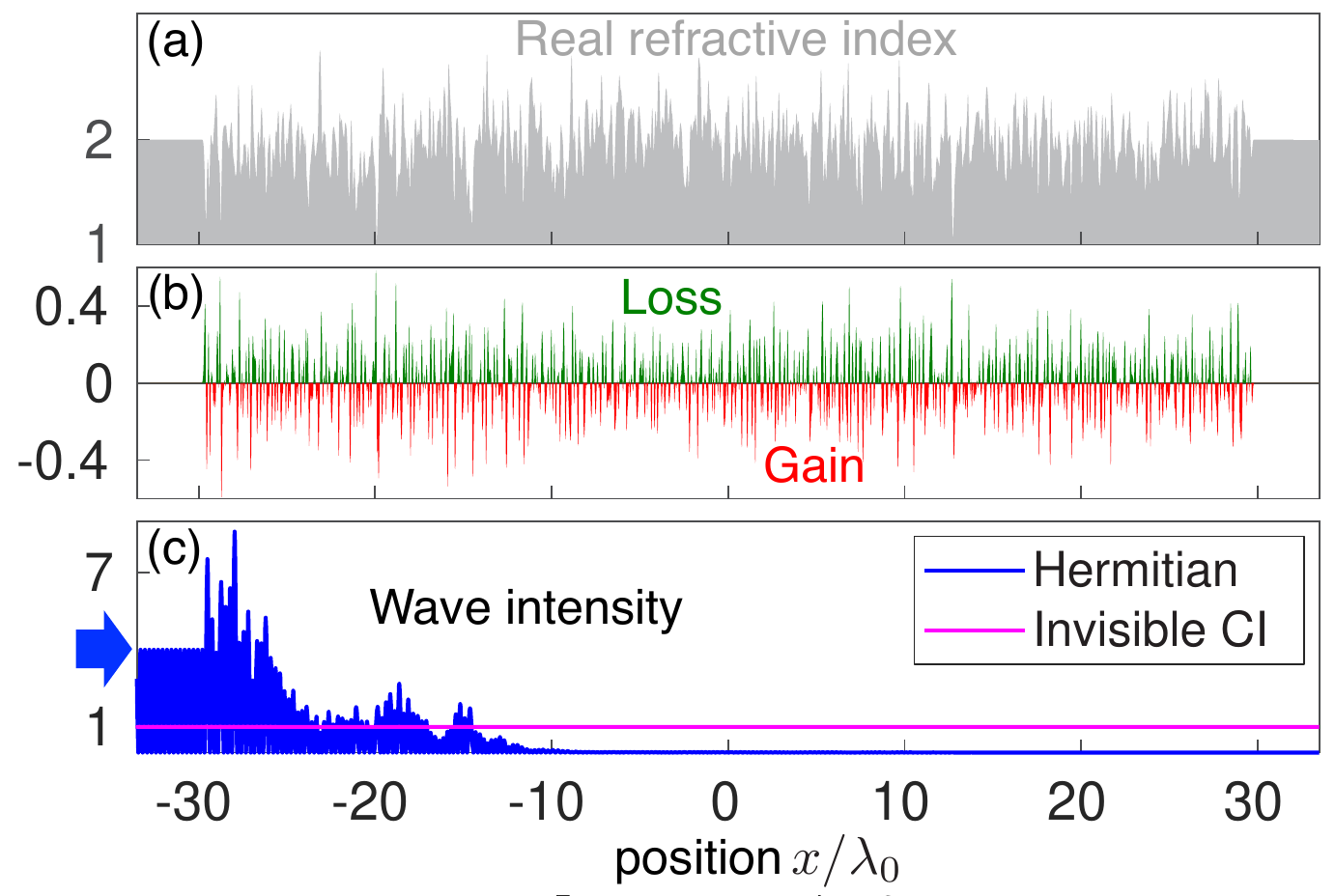}
\caption{(a), (b) Real (gray) and imaginary part (red: gain, green: loss) of the refractive index for a strongly disordered CI system. We use Eq.~(\ref{eq:CI_ref_index}) with $W(x) = n_0+ f(x)$, where $f(x)$ is a superposition of $N=3000$ Gaussian functions with different widths ($\sigma$ uniformly distributed between $0.04\lambda_0$ and $0.05\lambda_0$), heights (uniformly distributed between $0$ and $0.22$) and positions, satisfying the invisibility condition in Eq.~(\ref{eq:invisibility_condition}). The asymptotic refractive index is $n_0=2$ and the wavenumber $k_0=2\pi/0.2$. (c) Intensity of the scattering state for a plane wave with wavenumber $k_0$ and amplitude $A=1$ injected from the left into the system for the structure including gain and loss (magenta line) and without gain and loss of (blue line). As can be seen, the non-Hermitian components make the wave perfectly transmitting and free of any intensity variations.}\label{fig:3}
\end{figure} 
\begin{figure}[t]
\includegraphics[clip,width=1\linewidth]{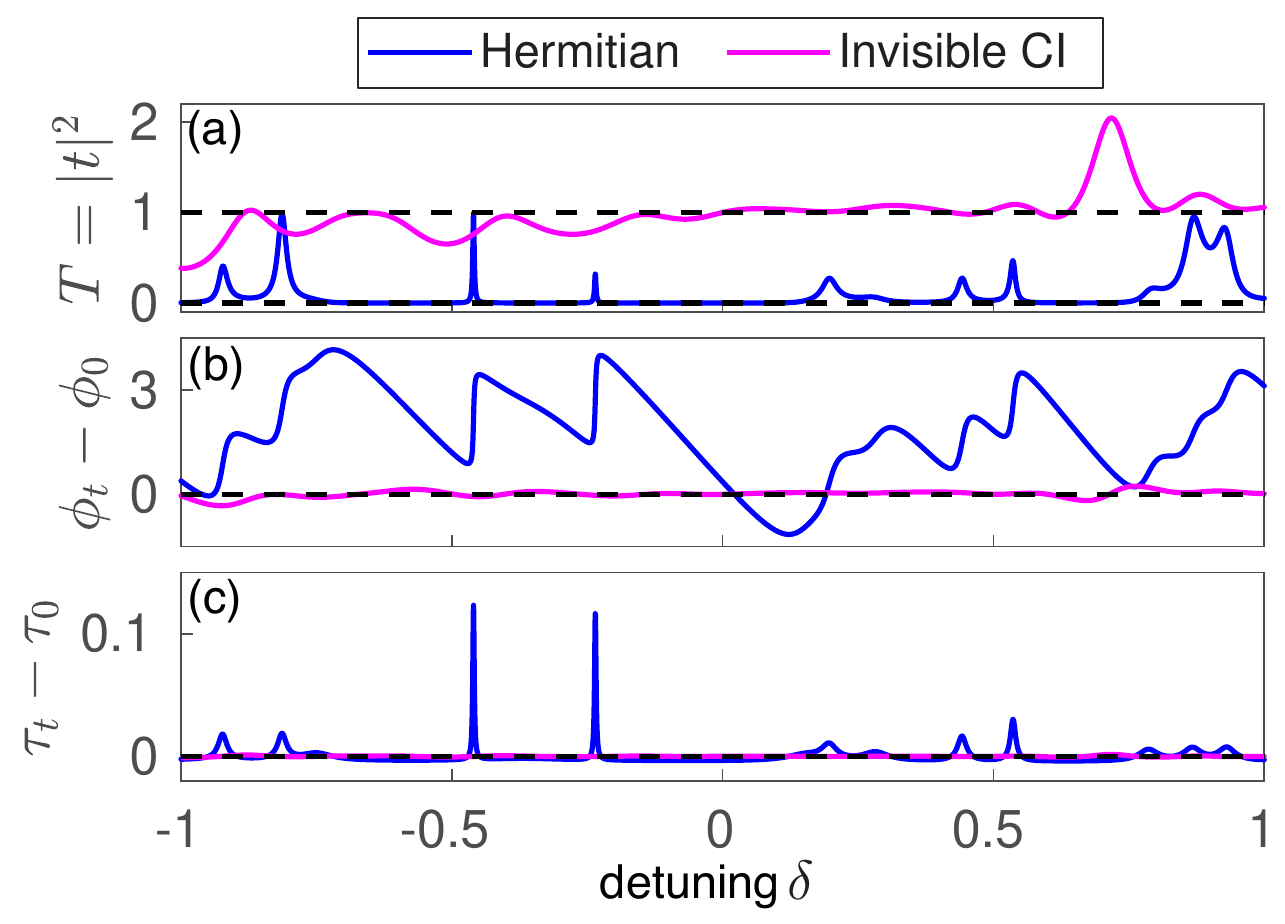}
\caption{(a) Transmittance, (b) difference in the transmission phase and (c) difference in the time-delay for the Hermitian (blue line) and for the invisible CI (magenta line) system shown in Fig.~\ref{fig:3} as a function of the detuning $\delta=k-k_0$. All quantities indicate that even such a disordered system is invisible for left-incident waves in a broad frequency range between $\delta=-0.5$ and $\delta=0.5$. The deviations of the invisible CI system from perfect invisibility (see horizontal dashed lines) are around two orders of magnitude smaller than for the Hermitian system (see appendix \ref{AppA}). The relative width of the invisibility window is $\Delta \delta / k_0 \approx 1/31.42 \approx 0.03$, i.e., a wavenumber detuning of around $1.5\%$ from $k_0$ in both directions still allows for perfect transmission and zero accumulated phase, which is significantly broader than a resonance in the Hermitian system.}\label{fig:4}
\end{figure} 

While this simple system already provides a first example to illustrate our protocol, we will now demonstrate its general applicability. Our starting point will be a disordered system whose strong variations in the refractive index lead to Anderson localization. The corresponding index profile follows the generating function in Eq.~(\ref{eq:generating_function}) with $f(x)$ being a superposition of $N=3000$ Gaussians with random width, height and position, satisfying the invisibility condition in Eq.~(\ref{eq:invisibility_condition}). The choice of using partially overlapping Gaussians is just for convenience here - any other arbitrary but smooth function $f(x)$ satisfying Eq.~(\ref{eq:invisibility_condition}) can also be used. In analogy to Fig.~\ref{fig:1}, we display the real and imaginary parts of the refractive index in Fig.~\ref{fig:3}(a) and (b), respectively, and the corresponding scattering states in (c). In appendix \ref{AppB} we calculate the localization length of the Hermitian system in Fig.~\ref{fig:3}(a), which turns out to be $\xi \approx 20 \lambda_0$, i.e., the system has a width of approximately $3\xi$ and is thus deeply in the localized regime. As a consequence, the wave gets hardly transmitted (blue line) in the Hermitian system whereas in the non-Hermitian case (magenta line) its transmission is perfect featuring constant intensity. We show now that even such a strongly disordered system is unidirectionally invisible. As displayed in Fig.~\ref{fig:4}, the CI system yields not only the same transmittance [see (a)], but also the same transmission phase [see (b)] and the same time-delay [see (c)] as the corresponding uniform system. We may thus conclude that the disordered structure in Fig.~\ref{fig:3}, which, in the absence of gain and loss is in the regime of Anderson localization, can be made completely invisible by adding the correct gain-loss refractive index distribution to it.

\begin{figure}[t]
\includegraphics[clip,width=1\linewidth]{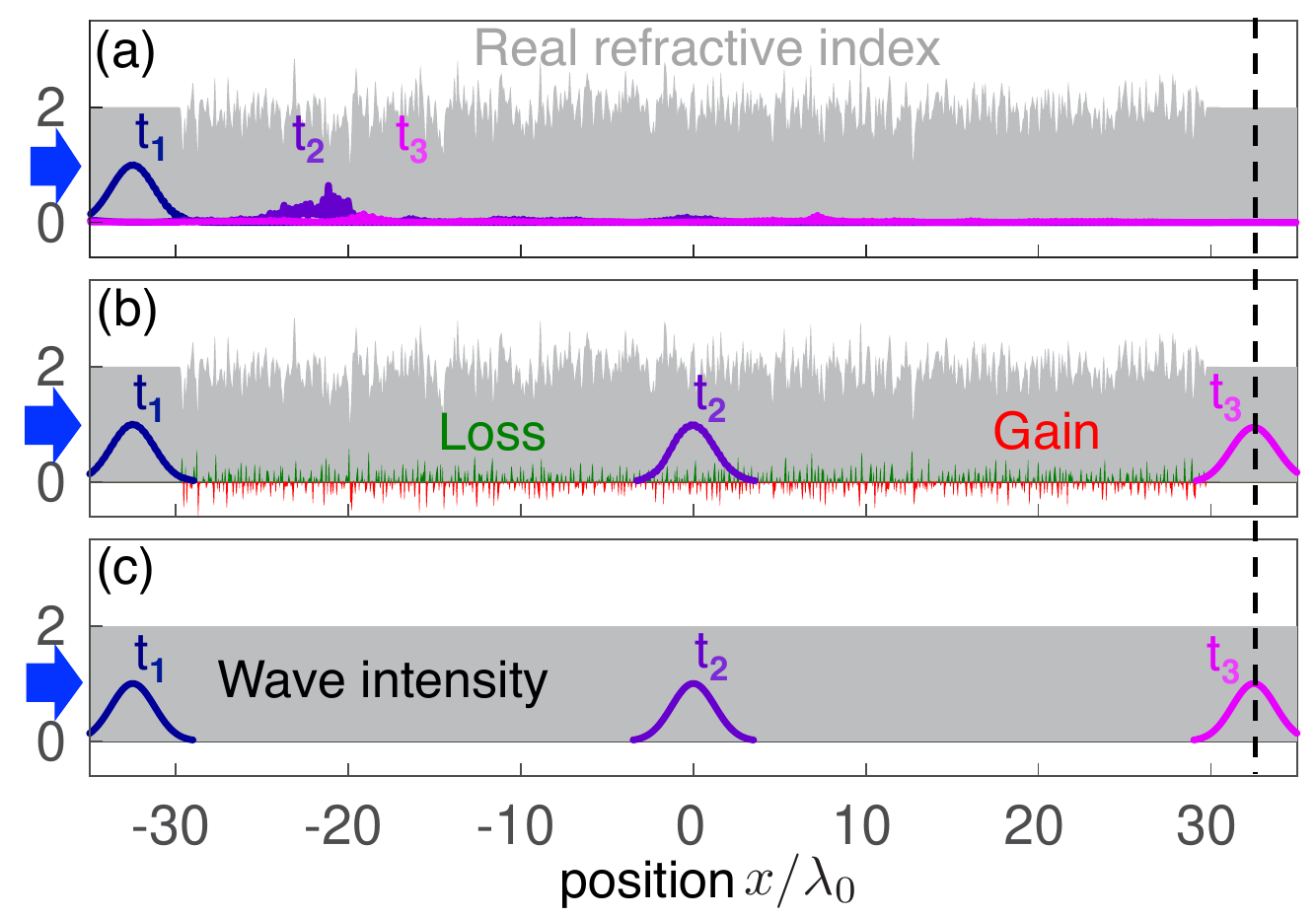}
\caption{Pulse propagating through (a) the disordered Hermitian system from Fig.~\ref{fig:3}, (b) the corresponding invisible CI system and (c) a uniform system at three different time steps $t_1<t_2<t_3$. Whereas the pulse gets reflected almost entirely at the Hermitian structure [see (a)], it gets perfectly transmitted through the CI system [see (b)] while maintaining its initial shape. Moreover, the pulse takes the same time to propagate through the structure as through a uniform system [see (c)]. The Fourier spectrum of the pulse is Gaussian-shaped with a standard deviation of $\sigma \approx 0.06k_0$.}\label{fig:5}
\end{figure}

The absence of any intensity variations in CI waves is due to the absence of back-reflections even inside the disordered medium. As such, CI waves are not a resonance phenomenon with a sharp frequency dependence, but they depend, instead, only weakly on frequency detuning, see Fig.~\ref{fig:4}. We will now make use of this broadband stability to test whether we can even launch pulses through our unidirectionally invisible potentials that feature the same time-delay as a pulse propagating through the corresponding uniform structure. We first show in Fig.~\ref{fig:5}(a) the propagation of a pulse through the same disordered Hermitian system as in Fig.~\ref{fig:3} at three different time-steps ($t_1<t_2<t_3$). As expected, for the Hermitian system with Anderson localization the pulse diffracts already considerably before reaching the other end of the structure. In stark contrast, we observe that  the pulse in the corresponding invisible CI system [see Fig.~\ref{fig:5}(b)] propagates through the system while maintaining its initial shape throughout the entire transmission process. Comparing this situation to a pulse propagating through a uniform system with the asymptotic index value $n_0=2$ [see Fig.~\ref{fig:5}(c)], we see that both pulses arrive at the end of the structure in the same shape and at the same time (as indicated by the vertical dashed line). Adding the appropriate gain-loss distribution to a disordered structure thus allows us not only to make the system perfectly transmitting for pulses but even completely invisible for them. 

This new approach for designing invisible structures that require neither periodicity nor any other symmetry (like $\mathcal{PT}$-symmetry) provides a significant step forward as compared to the concept presented in Ref.~\cite{lin_unidirectional_2011}. The $\mathcal{PT}$-symmetric Bragg grating presented there with $n(x)=n_0+n_1\cos(2\beta x)+\mathrm{i}n_2\sin(2\beta x)$ and $\beta$ being the spatial frequency of the grating, $n_0$ the asymptotic refractive index and $n_1$ and $n_2$ the amplitudes of the index variations, was shown to be unidirectionally invisible around the Bragg point ($\beta = k_0$) and with $n_1=n_2$. The asymptotic refractive index was assumed to be $n_0=1$ and the index variations were small $n_1=n_2=10^{-3}$. We show in appendix \ref{AppC} that this Bragg structure with the above listed parameters coincides with an invisible CI system derived from the generating function $W(x)= n_0+n'\cos(2\beta' x)$, if (i) the system is at the Bragg point $\beta' = k_0$ and (ii) the index variations are small, $n'=10^{-3}$. This finding explains why in an experimental realization of such a Bragg grating with loss elements only \cite{feng_experimental_2013} the wave intensity was found to be a pure exponential decay, which is the counterpart of a constant intensity in effective $\mathcal{PT}$-symmetric systems without gain \cite{guo_observation_2009}. When moving away from the Bragg point ($\beta \ne k_0$) and considering larger index variations ($n_1=n_2\approx 10^{-1}$), the Bragg grating from Ref.~\cite{lin_unidirectional_2011} is no longer unidirectionally invisible, whereas the corresponding invisible CI system maintains its invisibility even away from the Bragg point and also for arbitrarily large index variations $n'$ (see appendix \ref{AppC}).

In summary, we show that the interplay of gain and loss allows for a new class of unidirectionally invisible systems that are very robust with respect to frequency variations without satisfying any spatial symmetries. Our approach constitutes a broadly applicable generalization of earlier concepts restricted to periodic, layered or analytic (in one half of the complex position plane) potentials. Even disordered systems, which, in the absence of gain and loss, give rise to Anderson localization, can be made unidirectionally invisible using our approach. The key concept to arrive at these results is that of constant-intensity waves, whose frequency stability even allows us to create pulses that propagate through disorder as through a uniform system. We are confident that these exciting predictions can be implemented in a number of experiments where the spatial engineering of gain and loss has recently been achieved successfully \cite{regensburger_parity-time_2012, ruter_observation_2010, feng_experimental_2013, feng_demonstration_2014, rivet_constant-pressure_2018}.

We acknowledge support by the European Commission under project NHQWAVE (MSCA-RISE 691209). AB is a recipient of a DOC Fellowship of the Austrian Academy of Sciences at the Institute of Theoretical Physics of Vienna University of Technology (TU Wien). The computational results presented have been partly achieved using the Vienna Scientific Cluster (VSC). We would also like to thank Ziad H. Musslimani for fruitful discussions. 

\begin{appendix}

\section{Deviation from perfect invisibility \label{AppA}}
In order to quantify how close the invisible CI systems shown in Fig.~\ref{fig:2} and Fig.~\ref{fig:4} of the main text are to perfect invisibility, we evaluate the following three quantities within the invisibility window (as defined in the caption of Fig.~\ref{fig:2} and Fig.~\ref{fig:4}): (i) the mean squared deviation (MSD) of the transmittance from perfect transmission ($T=1$), (ii) the mean squared deviation of the phase difference from zero extra phase ($\phi_t-\phi_0=0$) and (iii) the mean squared deviation of the time-delay difference from zero extra time-delay ($\tau_t-\tau_0=0$). In order to get reference values, we evaluate all deviations for the corresponding Hermitian systems as well. For the system shown in Fig.~\ref{fig:2}, we get the following results: 
\begin{itemize}
\item $MSD(T)_{Herm} \approx 5.01\cdot 10^{-2}$

 $MSD(T)_{CI}\approx 6.9\cdot 10^{-4}$
\item $MSD(\phi_t-\phi_0)_{Herm} \approx 6.72\cdot 10^{-3}$

$MSD(\phi_t-\phi_0)_{CI} \approx 1.98\cdot 10^{-5}$

\item $MSD(\tau_t-\tau_0)_{Herm}\approx 1.34 \cdot 10^{-6}$

$MSD(\tau_t-\tau_0)_{CI}\approx 1.37\cdot 10^{-8}$
\end{itemize}
For the disordered system shown in Fig.~\ref{fig:4}, we obtain:

\begin{itemize}
\item $MSD(T)_{Herm} \approx 9.43\cdot 10^{-1}$

 $MSD(T)_{CI}\approx 1.29\cdot 10^{-2}$
\item $MSD(\phi_t-\phi_0)_{Herm} \approx 3.94$

$MSD(\phi_t-\phi_0)_{CI} \approx 1.9\cdot 10^{-3}$

\item $MSD(\tau_t-\tau_0)_{Herm}\approx 8.32 \cdot 10^{-5}$

$MSD(\tau_t-\tau_0)_{CI}\approx 2.94\cdot 10^{-8}$
\end{itemize}

These results clearly demonstrate that the deviations of our invisible CI systems from perfect invisibility are negligibly small and around two orders of magnitude smaller than for the corresponding Hermitian system. 

\section{Localization length of disordered structure \label{AppB}}
\begin{figure}[t]
\includegraphics[clip,width=1\linewidth]{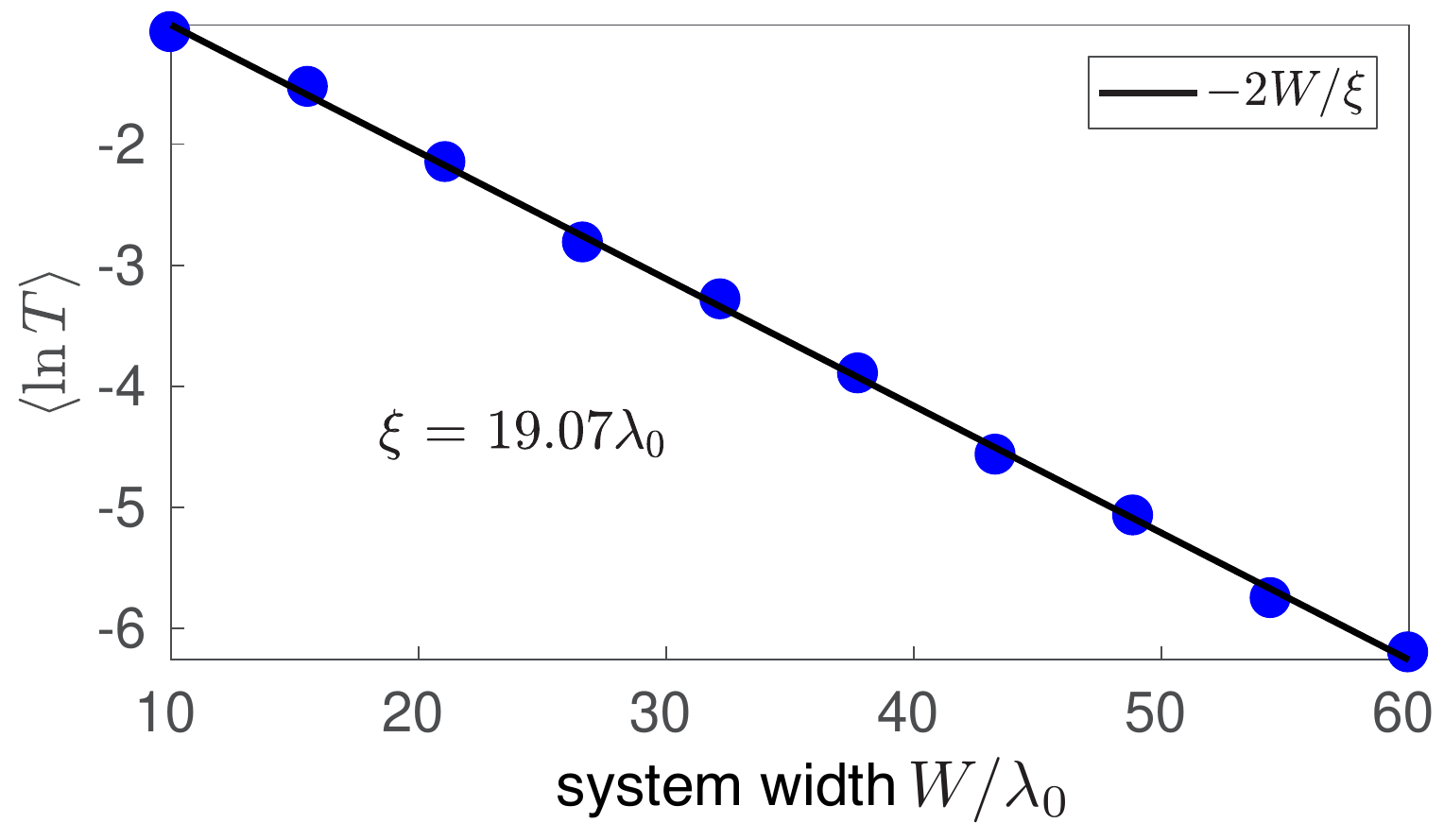}
\caption{Logarithmic transmittance averaged over $1000$ random configurations of the system shown in Fig.~\ref{fig:3}(a) as a function of the system's width $W=2L$. We fit the data to the black line $-2W/\xi$ whose slope we can use to estimate the localization length $\xi \approx 20\lambda_0$.} \label{fig:supp1}
\end{figure}

Here we show that the disordered structure in Fig.~\ref{fig:3}(a) gives rise to Anderson localization in the absence of gain and loss. To prove this explicitly, we determine its localization length $\xi$ which quantifies the exponential decrease of the transmittance $T=|t|^2$ as a function of the system's width $W=2L$. To be more precise, the localization length $\xi$ can be estimated by $\xi=-2W\langle \ln [T(W)] \rangle ^{-1}$, where the brackets $\langle \ldots \rangle$ denote the average value over $1000$ random configurations at a given system width $W$. In Fig.~\ref{fig:supp1} we plot the quantity $\langle \ln [T(W)] \rangle$ as a function of $W$ from which we can estimate the localization length $\xi$ through a fit with $-2W/\xi$ (black line). We find that the localization length is $\xi\approx 20 \lambda_0$, i.e., the disordered structure in Fig.~\ref{fig:3}(a) is around three times wider than the localization length $\xi$ and therefore deep in the localized regime. 

\section{Connection to unidirectionally invisible Bragg grating \label{AppC}}

The unidirectionally invisible Bragg grating introduced in Ref.~\cite{lin_unidirectional_2011} follows a $\mathcal{PT}$-symmetric and periodic refractive index modulation: 
\begin{equation}
\label{eq:zin_lin_inv}
n(x)=n_0+n_1\cos(2\beta x)+\mathrm{i}n_2\sin(2\beta x),
\end{equation}
with $\beta$ being the spatial periodicity of the grating. At the Bragg point ($\beta=k_0$) and with $n_1=n_2=10^{-3}$ and $n_0=1$, the structure becomes unidirectionally invisible for left incident waves and strongly reflecting for waves incident from the right. Whereas in \cite{lin_unidirectional_2011} this phenomenon of unidirectional invisibility is directly associated with the $\mathcal{PT}$-nature of this periodic structure, we show now that the structure in Eq.~(\ref{eq:zin_lin_inv}) coincides with  one example of our new class of unidirectionally invisible systems at one specific parameter configuration. We start with a generating function $W(x)= n_0  +f(x)$ featuring $f(x)=n' \cos(2\beta' x)$, which satisfies the invisibility condition in Eq.~(\ref{eq:invisibility_condition}), and calculate the corresponding CI refractive index from Eq.~(\ref{eq:CI_ref_index}):
\begin{align}
\label{eq:CI_cos}
n(x) \nonumber &= \sqrt{W^2(x)-\frac{\mathrm{i}}{k_0}\frac{dW(x)}{dx}} \nonumber \\ 
&=n_0\sqrt{1+\frac{n'^2}{n_0^2}\cos^2(2\beta' x)} \nonumber \\
& \indent + \overline{\frac{2n'}{n_0}\cos(2\beta' x)+\mathrm{i}\frac{2\beta' n'}{k_0 n_0^2}\sin(2\beta' x)}.
\end{align}
Assuming that $n'=10^{-3}$ (in analogy to Ref.~\cite{lin_unidirectional_2011}), we can neglect in Eq.~(\ref{eq:CI_cos}) the term which is proportional to $n'^2$ and consider the other two terms, which are proportional to $n'$, as small, allowing us to approximate the square root $\sqrt{1+x}\approx 1 + x/2$ for small $x$ [with $x=\frac{2n'}{n_0}\cos(2\beta' x)+\mathrm{i}\frac{2\beta' n'}{k_0 n_0^2}\sin(2\beta' x)$]. We end up with $n(x) \approx n_0+n'\cos(2\beta' x)+\mathrm{i}\frac{\beta' n'}{k_0 n_0}\sin(2\beta' x)$, which turns out to match exactly the structure in Eq.~(\ref{eq:zin_lin_inv}) if $n'=n_1=n_2=10^{-3}$, $n_0=1$ and $\beta'=\beta=k_0$. Our findings thus strongly indicate that the unidirectionally invisible Bragg grating in Ref.~\cite{lin_unidirectional_2011} is in fact a refractive index that supports CI waves which also satisfies the invisibility condition in Eq.~(\ref{eq:invisibility_condition}).

\begin{figure}[h!]
\includegraphics[clip,width=1\linewidth]{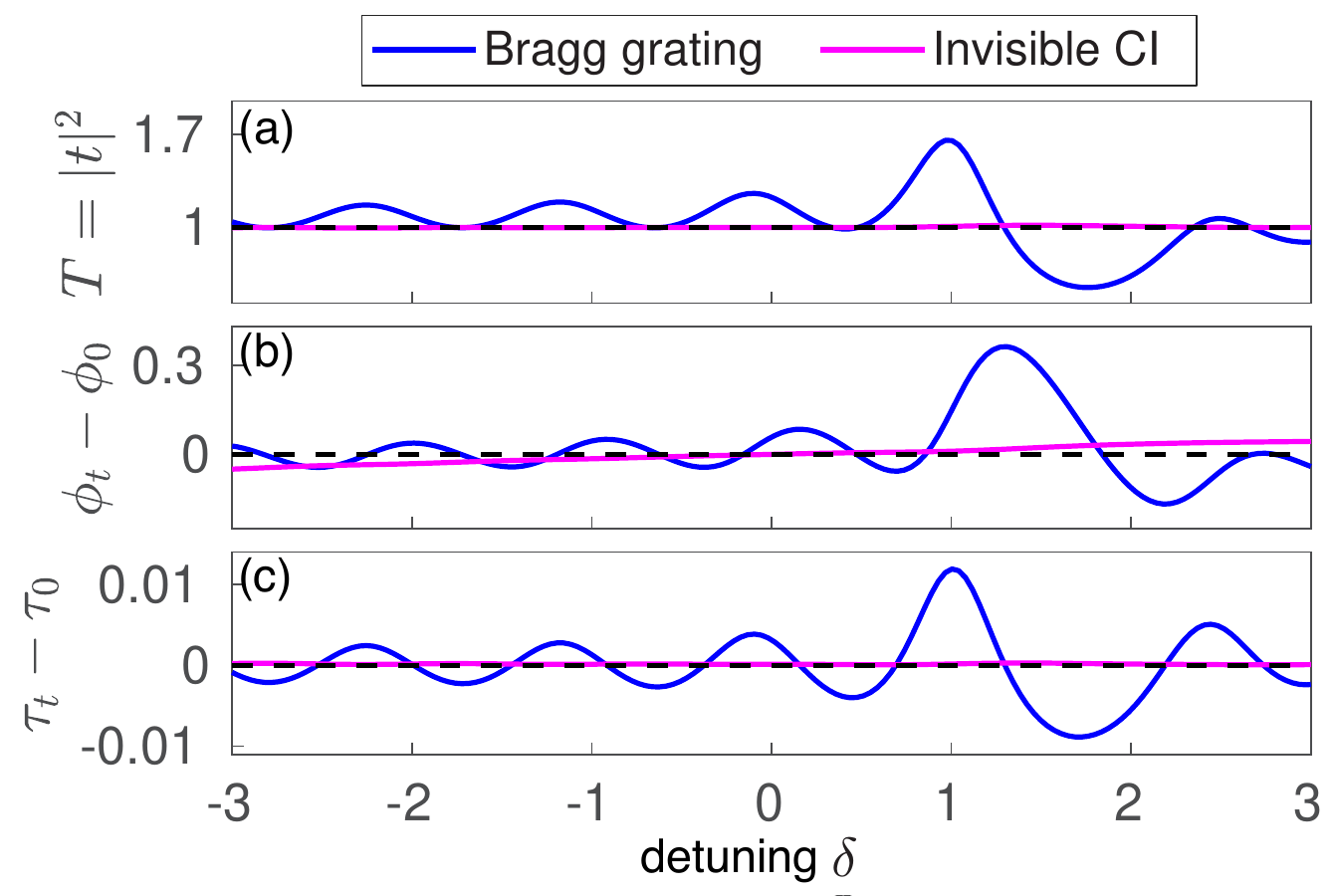}
\caption{(a) Transmittance, (b) difference in the transmission phase and (c) difference in the time-delay for the unidirectionally invisible Bragg structure in [Eq.~(\ref{eq:zin_lin_inv})] (blue line) with different parameters for $n_0, n_1, n_2, \beta$ as in Ref.~\cite{lin_unidirectional_2011} and for the CI system [Eq.~(\ref{eq:CI_cos})] (magenta line) as a function of the detuning $\delta = k-k_0$, with the following parameters: $n_0=2$, $n'=n_1=n_2=0.5$, $\beta'=\beta = 0.7 k_0$ and $k_0=2\pi/0.2$. The system's width is $\approx 7 \lambda_0$. All quantities indicate that the CI system is unidirectionally invisible in a broad frequency window, whereas the Bragg grating from Eq.~(\ref{eq:zin_lin_inv}) can already be detected by measuring the (frequency dependent) transmittance. The relative width of the invisibility window between $\delta=-2$ and $\delta = 2$ is $\Delta \delta / k_0 \approx 4/31.42 \approx 0.13$, i.e., a wavenumber detuning of around $6\%$ from $k_0$ in both directions still allows for perfect transmission and zero accumulated phase.}\label{fig:supp2}
\end{figure}

What makes the CI structure in Eq.~(\ref{eq:CI_cos}) superior, however, is the fact that it is invisible for all values of $\beta'$, $n_0$ and $n'$, i.e., it is neither restricted to the Bragg point $\beta' = k_0$ nor to small index variations $n'$. To prove our statement numerically, we perform the same calculations as in Fig.~\ref{fig:2} and Fig.~\ref{fig:4} of the main text, but now for two different systems: the first one [see Fig.~\ref{fig:supp2} blue lines] is the Bragg grating defined in Eq.~(\ref{eq:zin_lin_inv}) away from the Bragg point, $\beta=0.7k_0$, and for larger index variations $n_1=n_2=0.5$, whereas the second one [see Fig.~\ref{fig:supp2} magenta lines] is our invisible CI refractive index in Eq.~(\ref{eq:CI_cos}) with the same parameters ($\beta'=\beta$, $n'=n_1=n_2$). From Fig.~\ref{fig:supp2}(a) we can already see that the Bragg grating can be detected already by measuring the transmittance at different frequencies, whereas the CI system has unit transmittance in a broad frequency window. Also the transmission phase [see Fig.~\ref{fig:supp2}(b)] and the time-delay [see Fig.~\ref{fig:supp2}(c)] indicate that the CI system is invisible -- quite in contrast to the Bragg grating. The results shown in Fig.~\ref{fig:supp2} demonstrate that the CI system in Eq.~(\ref{eq:CI_cos}) is invisible per construction for arbitrary parameters $n_0$, $n'$ and $\beta'$, whereas the Bragg grating in Eq.~(\ref{eq:zin_lin_inv}) is only invisible for the parameters used in Ref.~\cite{lin_unidirectional_2011}. 

\end{appendix}

\bibliographystyle{apsrev4-1}

\bibliography{bibliography}
%\bibliography{scat_free_pulse}

\end{document}